\documentstyle[twoside,epsf,psfig,times]{article}
\def\beqa{\begin{eqnarray}}
\def\eeqa{\end{eqnarray}}
\def\beq{\begin{equation}}
\def\eeq{\end{equation}}
\catcode`\@=11
\long\def\@makefntext#1{
\protect\noindent \hbox to 3.2pt {\hskip-.9pt  
$^{{\eightrm\@thefnmark}}$\hfil}#1\hfill}               

\def\@makefnmark{\hbox to 0pt{$^{\@thefnmark}$\hss}}    

\def\ps@myheadings{\let\@mkboth\@gobbletwo
\def\@oddhead{\hbox{}
\rightmark\hfil\eightrm\thepage}   
\def\@oddfoot{}\def\@evenhead{\eightrm\thepage\hfil
\leftmark\hbox{}}\def\@evenfoot{}
\def\sectionmark##1{}\def\subsectionmark##1{}}

\oddsidemargin=\evensidemargin
\newcounter{sectionc}\newcounter{subsectionc}\newcounter{subsubsectionc}
\renewcommand{\section}[1] {\vspace{12pt}\addtocounter{sectionc}{1} 
\setcounter{subsectionc}{0}\setcounter{subsubsectionc}{0}\noindent 
        {\tenbf\thesectionc. #1}\par\vspace{5pt}}
\renewcommand{\subsection}[1] {\vspace{12pt}\addtocounter{subsectionc}{1} 
\setcounter{subsubsectionc}{0}\noindent 
{\bf\thesectionc.\thesubsectionc. {\kern1pt \bfit #1}}\par\vspace{5pt}}
\renewcommand{\subsubsection}[1] {\vspace{12pt}\addtocounter{subsubsectionc}{1}
        \noindent{\tenrm\thesectionc.\thesubsectionc.\thesubsubsectionc.
        {\kern1pt \tenit #1}}\par\vspace{5pt}}
\newcommand{\nonumsection}[1] {\vspace{12pt}\noindent{\tenbf #1}
        \par\vspace{5pt}}

\newcounter{appendixc}
\newcounter{subappendixc}[appendixc]
\newcounter{subsubappendixc}[subappendixc]
\renewcommand{\thesubappendixc}{\Alph{appendixc}.\arabic{subappendixc}}
\renewcommand{\thesubsubappendixc}
        {\Alph{appendixc}.\arabic{subappendixc}.\arabic{subsubappendixc}}

\renewcommand{\appendix}[1] {\vspace{12pt}
        \refstepcounter{appendixc}
        \setcounter{figure}{0}
        \setcounter{table}{0}
        \setcounter{lemma}{0}
        \setcounter{theorem}{0}
        \setcounter{corollary}{0}
        \setcounter{definition}{0}
        \setcounter{equation}{0}
        \renewcommand{\thefigure}{\Alph{appendixc}.\arabic{figure}}
        \renewcommand{\thetable}{\Alph{appendixc}.\arabic{table}}
        \renewcommand{\theappendixc}{\Alph{appendixc}}
        \renewcommand{\thelemma}{\Alph{appendixc}.\arabic{lemma}}
        \renewcommand{\thetheorem}{\Alph{appendixc}.\arabic{theorem}}
        \renewcommand{\thedefinition}{\Alph{appendixc}.\arabic{definition}}
        \renewcommand{\thecorollary}{\Alph{appendixc}.\arabic{corollary}}
        \renewcommand{\theequation}{\Alph{appendixc}.\arabic{equation}}
        \noindent{\tenbf Appendix \theappendixc #1}\par\vspace{5pt}}
\newcommand{\subappendix}[1] {\vspace{12pt}
        \refstepcounter{subappendixc}
        \noindent{\bf Appendix \thesubappendixc. {\kern1pt \bfit #1}}
        \par\vspace{5pt}}
\newcommand{\subsubappendix}[1] {\vspace{12pt}
        \refstepcounter{subsubappendixc}
        \noindent{\rm Appendix \thesubsubappendixc. {\kern1pt \tenit #1}}
        \par\vspace{5pt}}

\newcommand{\textlineskip}{\baselineskip=13pt}
\newcommand{\smalllineskip}{\baselineskip=10pt}

\newcommand{\copyrightheading}[1]
        {\vspace*{-2.5cm}\smalllineskip{\flushleft
        {\footnotesize International Journal of Modern Physics D, #1}\\
        {\footnotesize \copyright\kern2pt World Scientific Publishing
         Company}\\
         }}

\newcommand{\publisher}[2]{{\begin{center}\footnotesize\smalllineskip 
        Received #1\\
        Revised #2
        \end{center}
        }}

\def\abstracts#1#2#3{{
        \centering{\begin{minipage}{4.5in}\footnotesize\baselineskip=10pt
        \parindent=0pt #1\par 
        \parindent=15pt #2\par
        \parindent=15pt #3
        \end{minipage}}\par}} 

\def\keywords#1{{
        \centering{\begin{minipage}{4.5in}\footnotesize\baselineskip=10pt
        {\footnotesize\it Keywords}\/: #1
         \end{minipage}}\par}}

\renewenvironment{thebibliography}[1]
        {\frenchspacing
         \ninerm\baselineskip=11pt
         \begin{list}{\arabic{enumi}.}
        {\usecounter{enumi}\setlength{\parsep}{0pt}     
         \setlength{\leftmargin 12.7pt}{\rightmargin 0pt}
         \setlength{\itemsep}{0pt} \settowidth
        {\labelwidth}{#1.}\sloppy}}{\end{list}}

\newcounter{itemlistc}
\newcounter{romanlistc}
\newcounter{alphlistc}
\newcounter{arabiclistc}

\newcommand{\fcaption}[1]{
        \refstepcounter{figure}
        \setbox\@tempboxa = \hbox{\footnotesize Fig.~\thefigure. #1}
        \ifdim \wd\@tempboxa > 5in
           {\begin{center}
        \parbox{5in}{\footnotesize\smalllineskip Fig.~\thefigure. #1}
            \end{center}}
        \else
             {\begin{center}
             {\footnotesize Fig.~\thefigure. #1}
              \end{center}}
        \fi}

\newcommand{\tcaption}[1]{
        \refstepcounter{table}
        \setbox\@tempboxa = \hbox{\footnotesize Table~\thetable. #1}
        \ifdim \wd\@tempboxa > 5in
           {\begin{center}
        \parbox{5in}{\footnotesize\smalllineskip Table~\thetable. #1}
            \end{center}}
        \else
             {\begin{center}
             {\footnotesize Table~\thetable. #1}
              \end{center}}
        \fi}

\def\@citex[#1]#2{\if@filesw\immediate\write\@auxout
        {\string\citation{#2}}\fi
\def\@citea{}\@cite{\@for\@citeb:=#2\do
        {\@citea\def\@citea{,}\@ifundefined
        {b@\@citeb}{{\bf ?}\@warning
        {Citation `\@citeb' on page \thepage \space undefined}}
        {\csname b@\@citeb\endcsname}}}{#1}}

\newif\if@cghi
\def\cite{\@cghitrue\@ifnextchar [{\@tempswatrue
        \@citex}{\@tempswafalse\@citex[]}}
\def\citelow{\@cghifalse\@ifnextchar [{\@tempswatrue
        \@citex}{\@tempswafalse\@citex[]}}
\def\@cite#1#2{{$\null^{#1}$\if@tempswa\typeout
        {IJCGA warning: optional citation argument 
        ignored: `#2'} \fi}}

\def\pmb#1{\setbox0=\hbox{#1}
        \kern-.025em\copy0\kern-\wd0
        \kern.05em\copy0\kern-\wd0
        \kern-.025em\raise.0433em\box0}

\def\fnt#1#2{\footnotetext{\kern-.3em
        {$^{\mbox{\scriptsize #1}}$}{#2}}}

\def\fpage#1{\begingroup
\voffset=.3in
\thispagestyle{empty}\begin{table}[b]\centerline{\footnotesize #1}
        \end{table}\endgroup}

\def\runninghead#1#2{\pagestyle{myheadings}
\markboth{{\protect\footnotesize\it{\quad #1}}\hfill}
{\hfill{\protect\footnotesize\it{#2\quad}}}}
\font\tenrm=cmr10
\font\tenit=cmti10 
\font\tenbf=cmbx10
\font\bfit=cmbxti10 at 10pt
\font\ninerm=cmr9

\font\eightrm=cmr8

\def\qed{\hbox{${\vcenter{\vbox{                  
   \hrule height 0.4pt\hbox{\vrule width 0.4pt height 6pt
   \kern5pt\vrule width 0.4pt}\hrule height 0.4pt}}}$}}

\begin{document}
\setlength{\textheight}{7.7truein}    

\runninghead{Some inhomogeneous magnetized viscous fluid cosmological models with
varying $\Lambda$} 
{A. Pradhan, S. K. Srivastva and K. R. Jotania}

\normalsize\textlineskip
\thispagestyle{empty}
\setcounter{page}{1}

\copyrightheading{}             {Vol.~0, No.~0 (2003) 000--000}

\vspace*{0.88truein}

\fpage{1}

\centerline{\bf SOME INHOMOGENEOUS  MAGNETIZED VISCOUS FLUID  COSMOLOGICAL}
\vspace*{0.035truein}
\centerline{\bf MODELS WITH VARYING $\Lambda$}
\vspace*{0.37truein}
\centerline{\footnotesize ANIRUDH PRADHAN\footnote{E-mail: acpradhan@yahoo.com,
pradhan@iucaa.ernet.in (Corresponding Author)}~ ~ SUDHIR KUMAR SRIVASTAV }
\vspace*{0.015truein}  
\centerline{\footnotesize\it Department of Mathematics, Hindu Post-graduate College,}
\baselineskip=10pt
\centerline{\footnotesize\it Zamania, Ghazipur 232 331, India}
\vspace*{10pt}
\centerline{\footnotesize KANTI R. JOTANIA\footnote{E-mail: kanti@iucaa.
ernet.in} }
\vspace*{0.015truein}
\centerline{\footnotesize\it Department of Physics, M. N. College, Visnagar
 384 315, Gujarat, India }
\baselineskip=10pt
\vspace*{0.225truein}
\publisher{(received date)}{(revised date)}
\vspace*{0.21truein}
\abstracts{Some cylindrically symmetric inhomogeneous viscous fluid cosmological 
models with electro-magnetic field are obtained. To get a solution a supplementary 
condition between metric potentials is used. The viscosity coefficient of bulk 
viscous fluid is assumed to be a power function of mass density. Without assuming 
any {\it ad hoc} law, we obtain a cosmological constant as a decreasing function 
of time. The behaviour of the electro-magnetic field tensor together with some 
physical aspects of the model are also discussed.}{}{}
\vspace*{10pt}


\vspace*{1pt}\textlineskip      
\section{Introduction}
\vspace*{-0.5pt}
Inhomogeneous cosmological models play an important role in understanding some 
essential features of the universe such as the formation of galaxies during the 
early stages of evolution and process of homogenization. The early attempts at the 
construction of such models have done by Tolman\cite{ref1} and Bondi\cite{ref2} 
who considered spherically symmetric models. Inhomogeneous plane-symmetric 
models were considered by Taub\cite{ref3,ref4} and later by Tomimura,\cite{ref5}
 Szekeres,\cite{ref6} Collins and Szafron,\cite{ref7} Szafron and Collins.
\cite{ref8} Recently, Senovilla\cite{ref9} obtained 
a new class of exact solutions of Einstein's equation without big bang singularity, 
representing a cylindrically symmetric, inhomogeneous cosmological model filled with 
perfect fluid which is smooth and regular everywhere satisfying energy and causality 
conditions. Later, Ruis and Senovilla\cite{ref10} have separated out a fairly large 
class of singularity free models through a comprehensive study of general cylindrically
symmetric metric with separable function of $r$ and $t$ as metric coefficients.
Dadhich {\it et al.}\cite{ref11} have established a link between the FRW model and the 
singularity free family by deducing the latter through a natural and simple
inhomogenization and anisotropization of the former. Recently, Patel 
{\it et al.}\cite{ref12}
presented a general class of inhomogeneous cosmological models filled with 
non-thermalized perfect fluid by assuming that the background spacetime admits 
two space-likecommuting killing vectors and has separable metric coefficients.
 Bali and Tyagi\cite{ref13}obtained a plane-symmetric inhomogeneous cosmological 
models of perfect fluid distribution with electro-magnetic field. 
Recently, Pradhan {\it et al.}\cite{ref14} have investigated
a plane-symmetric inhomogeneous viscous fluid cosmological models with 
electro-magnetic field.     
\newline
\par
Models with a relic cosmological constant $\Lambda$ have received considerable 
attention recently among researchers for various reasons 
(see Refs.{\cite{ref15}}$^-${\cite{ref19}} and references therein). Some of the 
recent discussions on the cosmological constant ``problem'' and consequence on
cosmology with a time-varying cosmological constant by Ratra and Peebles,\cite{ref20} 
Dolgov\cite{ref21}$^-$\cite{ref23} and Sahni and Starobinsky\cite{ref24} have
pointed out that in the absence of any interaction with matter or radiation, the 
cosmological constant remains a ``constant''. However, in the presence of
interactions with matter or radiation, a solution of Einstein equations and the 
assumed equation of covariant conservation of stress-energy with a time-varying 
$\Lambda$ can be found. For these solutions, conservation of energy requires 
decrease in the energy density of the vacuum component to be compensated by a 
corresponding increase in the energy density of matter or radiation. Earlier 
researchers on this topic, are contained in Zeldovich,\cite{ref25} 
Weinberg\cite{ref16} and Carroll, Press and Turner.\cite{ref26} Recent
observations by Perlmutter {\it et al.}\cite{ref27} and Riess {\it et al.} \cite{ref28}
strongly favour a significant and positive value of $\Lambda$. Their finding 
arise from the study of more than $50$ type Ia supernovae with redshifts in the range
$0.10 \leq z \leq 0.83$ and these suggest Friedmann models with negative pressure
matter such as a cosmological constant $(\Lambda)$, domain walls or cosmic 
strings (Vilenkin,\cite{ref29} Garnavich {\it et al.}\cite{ref30}) 
Recently, Carmeli and Kuzmenko\cite{ref31}
have shown that the cosmological relativistic theory (Behar and Carmeli\cite{ref32})
predicts the value for cosmological constant $\Lambda = 1.934\times 10^{-35} s^{-2}$.
This value of ``$\Lambda$'' is in excellent agreement with the measurements 
recently obtained by the High-Z Supernova Team and Supernova Cosmological 
Project (Garnavich {\it et al.}\cite{ref30}; Perlmutter {\it et al.}
\cite{ref27}; Riess {\it et al.}\cite{ref28}; Schmidt {\it  et al.}\cite{ref31}) 
The main conclusion of these observations is that the expansion 
of the universe is accelerating. 
\newline
\par
Several ans$\ddot{a}$tz have been proposed in which the $\Lambda$ term decays 
with time (see Refs. Gasperini,\cite{ref34,ref35} Berman,\cite{ref36} 
Freese {\it et al.},\cite{ref19} $\ddot{O}$zer and Taha,\cite{ref19} 
Peebles and Ratra,\cite{ref37} Chen and Hu,\cite{ref38} Abdussattar and 
Viswakarma,\cite{ref39} Gariel and Le Denmat,\cite{ref40} Pradhan {\it et al.}
\cite{ref41}). Of the special interest 
is the ans$\ddot{a}$tz $\Lambda \propto S^{-2}$ (where $S$ is the scale factor of the
Robertson-Walker metric) by Chen and Wu,\cite{ref38} which has been 
considered/modified by several authors ( Abdel-Rahaman,\cite{ref42} 
Carvalho {\it et al.},\cite{ref19} Waga,\cite{ref43} Silveira and Waga,\cite{ref44}
Vishwakarma\cite{ref45}).
\newline
\par
Most cosmological models assume that the matter in the universe can be described 
by 'dust'(a pressure-less distribution) or at best a perfect fluid. 
However, bulk viscosity is expected to play an important role at certain stages
of expanding universe.\cite{ref46}$^-$\cite{ref48} It has been shown that bulk
viscosity leads to inflationary like solution,\cite{ref49} and acts like a negative
energy field in an expanding universe.\cite{ref50} Furthermore, there are several
processes which are expected to give rise to viscous effects. These are the decoupling
of neutrinos during the radiation era and the decoupling of radiation and matter 
during the recombination era. Bulk viscosity is associated with the Grand Unification 
Theories (GUT) phase transition and string creation. Thus, we should consider 
the presence of a material distribution other
than a perfect fluid to have realistic cosmological models (see Gr\o n\cite{ref51}
for a review on cosmological models with bulk viscosity). A number of authors 
have discussed cosmological solutions with bulk viscosity in various context.
\cite{ref51}$^-$\cite{ref54}
\newline
\par
The occurrence of magnetic fields on galactic scale is well-established 
fact today, and their importance for a variety of astrophysical phenomena
is generally acknowledged as pointed out Zeldovich {\it et al.}\cite{ref55} Also 
Harrison\cite{ref56} has suggested that magnetic field could have a 
cosmological origin. As a natural consequences, we should include magnetic 
fields in the energy-momentum tensor of the early universe.
The choice of anisotropic cosmological models in Einstein system of field 
equations leads to the cosmological models more general than Robertson-Walker 
model.\cite{ref57} The presence of primordial magnetic fields in the early 
stages of the evolution of the universe has been discussed by several authors.
\cite{ref58} $^-$\cite{ref67} Strong magnetic fields can be created due to 
adiabatic compression in clusters of galaxies. Large-scale magnetic fields 
give rise to anisotropies in the universe. The  anisotropic pressure created 
by the magnetic fields dominates the evolution of the shear anisotropy and it 
decays slower than if the pressure was isotropic.\cite{ref68,ref69} Such 
fields can be generated at the end of an inflationary epoch.\cite{ref70}$^-$
\cite{ref74} Anisotropic magnetic field models have significant contribution 
in the evolution of galaxies and stellar objects. Bali and Ali\cite{ref75} 
had obtained a magnetized cylindrically symmetric universe with an 
electrically neutral perfect fluid as the source of matter. Several authors
\cite{ref76}$^-$\cite{ref81} have investigated Bianchi type I cosmological 
models with a magnetic field in different context.\\
\par
Recently Singh {\it et al.}\cite{ref82} obtained some cylindrically symmetric 
inhomogeneous cosmological models for a perfect fluid distribution with 
electro-magnetic field. Motivated the situations discussed above, in this paper, 
we shall focus upon the problem of establishing a formalism for studying the 
general relativistic evolution magnetic inhomogeneities in presence of bulk 
viscous in an expanding universe. We do this by extending the work of Singh
{\it et al.}\cite{ref82} by including an electrically neutral bulk viscous fluid 
as the source of matter in the energy-momentum tensor. This paper is organised as 
follows. The metric and the field equations are presented in section 2. In section
 3 we deal with the  solution of the field equations in presence of bulk viscous 
fluid. The sections 3.1, 3.2 and 3.3  contain the three cases as $K > 0$, $K < 0$
and $K = 0$ and also contain some physical aspects of these models respectively. 
Finally in section 4 concluding remarks have been given. \\ 
\par
\section{The metric and field  equations}
We consider the metric in the form  
\begin{equation} 
\label{eq1}  
ds^{2} = A^{2}(dx^{2} - dt^{2}) + B^{2}dy^{2} + C^{2}dz^{2},
\end{equation} 
where the metric potential $A$ is a function of $t$ alone and  $B$ and $C$ are functions 
of $x$ and $t$. \\
The energy momentum tensor in the presence of bulk stress has the form
\begin{equation} 
\label{eq2}
T^{j}_{i} = (\rho + \bar{p})v_{i}v^{j} + \bar{p}g^{j}_{i} + E^{j}_{i},
\end{equation} 
where $E^{j}_{i}$ is the electro-magnetic field given by Lichnerowicz\cite{ref83}
as 
\begin{equation} 
\label{eq3}
E^{j}_{i} = \bar{\mu}\left[|h|^{2}\left(v_{i}v^{j} + \frac{1}{2}g^{j}_{i}\right)
- h_{i}h^{j}\right]
\end{equation} 
and
\begin{equation} 
\label{eq4}
\bar{p} = p - \xi v^{i}_{;i}
\end{equation} 
Here $\rho$, $p$, $\bar{p}$ and $\xi$ are the energy density, isotropic pressure
, effective pressure, bulk viscous coefficient respectively and $v^{i}$ is the 
flow vector satisfying the relation
\begin{equation} 
\label{eq5}
g_{ij} v^{i}v^{j} = - 1
\end{equation} 
$\bar\mu$ is the magnetic permeability and $h_{i}$ the magnetic flux vector
defined by
\begin{equation} 
\label{eq6}
h_{i} = \frac{1}{\bar{\mu}}~~ ^*F_{ji}v^{j}
\end{equation} 
where $^*F_{ij}$ is the dual electro-magnetic field tensor defined by Synge
\cite{ref84} to be
\begin{equation} 
\label{eq7}
^*F_{ij} = \frac{\sqrt-g}{2}\epsilon_{ijkl} F^{kl}
\end{equation} 
$F_{ij}$ is the electro-magnetic field tensor and $\epsilon_{ijkl}$ is the
Levi-Civita tensor density. Here, the comoving coordinates are taken to be 
$v^{1}$ = $0$ = $v^{2}$ = $v^{3}$ and $v^{4}$ = $\frac{1}{A}$. We consider the  
current to be flowing along the $z$-axis so that $F_{12}$ is the only non-
vanishing component of $F_{ij}$. The Maxwell's equations
\begin{equation} 
\label{eq8}
F_{[ij,k]} = 0,
\end{equation}  
\begin{equation} 
\label{eq9} 
\left[\frac{1}{\bar{\mu}}F^{ij}\right]_{;j} = J^{i},
\end{equation}  
require that $F_{12}$ is a function of $x$ alone. Here the semicolon represents a 
covariant differentiation. We assume that the magnetic permeability is 
a function of $x$ and $t$.   
The Einstein's field equations ( in gravitational units c = 1, G = 1 ) read as 
\begin{equation} 
\label{eq10} 
R^{j}_{i} - \frac{1}{2} R g^{j}_{i} + \Lambda g^{j}_{i} = - 8\pi T^{j}_{i},
\end{equation}  
for the line element (1) has been set up as
\[
8\pi A^{2}\left(\bar{p} + \frac{F^{2}_{12}}{2\bar{\mu}A^{2}B^{2}}\right) = 
\]
\begin{equation} 
\label{eq11}
\frac{A_{4} B_{4}}{AB} + \frac{A_{4}C_{4}}{AC} - \frac{B_{44}}{B} - \frac{C_{44}}{C} 
- \frac{B_{4}C_{4}}{BC} + \frac{B_{1}C_{1}}{BC} - \Lambda A^{2},
\end{equation}  
\begin{equation} 
\label{eq12} 
8\pi A^{2}\left(\bar{p} + \frac{F^{2}_{12}}{2\bar{\mu}A^{2}B^{2}}\right) =  
\frac{A^{2}_{4}}{A^{2}} - \frac{A_{44}}{A} - \frac{C_{44}}{C} + \frac{C_{11}}{C} 
- \Lambda A^{2},
\end{equation}  
\begin{equation} 
\label{eq13} 
8\pi A^{2}\left(\bar{p} - \frac{F^{2}_{12}}{2\bar{\mu}A^{2}B^{2}}\right) = 
\frac{A^{2}_{4}}{A^{2}} - \frac{A_{44}}{A} - \frac{B_{44}}{B} + \frac{B_{11}}{B} 
- \Lambda A^{2},
\end{equation}  
\[
8\pi A^{2}\left(\rho +  \frac{F^{2}_{12}}{2\bar{\mu}A^{2}B^{2}}\right) = 
\]
\begin{equation} 
\label{eq14} 
\frac{A_{4}B_{4}}{AB} + \frac{A_{4}C_{4}}{AC} - \frac{B_{11}}{B} - \frac{C_{11}}{C} 
- \frac{B_{1}C_{1}}{BC} + \frac{B_{4}C_{4}}{BC} - \Lambda A^{2},
\end{equation}  
\begin{equation} 
\label{eq15}
0 = \frac{B_{14}}{B} + \frac{C_{14}}{C} - \frac{A_{4}}{A}\left(\frac{B_{1}}{B} + 
\frac{C_{1}}{C}\right),
\end{equation}
where
\[
\bar{p} = p - \frac{\xi}{A}\left(\frac{A_{4}}{A} + \frac{B_{4}}{B} + \frac{C_{4}}{C}\right)
\]
The suffixes $1$ and $4$ by the symbols $A$, $B$ and $C$ denote differentiation with
respect to $x$ and $t$ respectively.\\
\par
\section{Solution of the field equations}
From Equations (\ref{eq11}), (\ref{eq12}) and (\ref{eq13}), we have
\begin{equation} 
\label{eq16}
\frac{A_{44}}{A} - \frac{A^{2}_{4}}{A^{2}} + \frac{A_{4}B_{4}}{AB}  + \frac{A_{4}C_{4}}{AC}
- \frac{B_{44}}{B} - \frac{B_{4}C_{4}}{BC} - \frac{C_{11}}{C} +  \frac{B_{1}C_{1}}{BC} = 0 
\end{equation} 
\begin{equation} 
\label{eq17}
\frac{8\pi F^{2}_{12}}{\bar\mu B^{2}} = - \frac{C_{44}}{C} + \frac{C_{11}}{C} +  
\frac{B_{44}}{B} - \frac{B_{11}}{B}
\end{equation}
Equations (\ref{eq11}) - (\ref{eq15}) represent a system of five equations in eight 
unknowns $A$, $B$, $C$, $\rho$, $p$, $F_{12}$, $\Lambda$ and $\xi$. To get a determinate 
solution, we need three extra conditions. \\
Firstly following Singh {\it et al.}\cite{ref81}, we assume a supplementary condition 
between the metric potentials as
\begin{equation} 
\label{eq18}
\frac{A_{4}}{A} = n \left(\frac{B_{4}}{B} - \frac{C_{4}}{C}\right),
\end{equation}
and
\begin{equation} 
\label{eq19}
\frac{C_{11}}{C} - \frac{B_{1}C_{1}}{BC} = K,
\end{equation}
where $n$ and $K$ are arbitrary constants.
Let us consider that
\[
B = f(x) g(t)
\]
\begin{equation} 
\label{eq20}
C = h(x)k(t)
\end{equation}
Using (\ref{eq18}) and (\ref{eq20}) in Equations (\ref{eq15}), (\ref{eq16}) and (\ref{eq17}),
we have
\begin{equation} 
\label{eq21}
\frac{(n + 1)\frac{\dot{k}}{k} - n\frac{\dot{g}}{g}}{(1 - n)\frac{\dot{g}}{g} + 
n \frac{\dot{k}}{k}} = - \frac{f^\prime/f}{h^\prime/h} = \alpha ~ ~ ({\rm constant, say}),
\end{equation}
\begin{equation} 
\label{eq22}
(n - 1)\frac{\ddot{g}}{g} - n \frac{\ddot{k}}{k} - \frac{\dot{g}\dot{k}}{gk} = K,
\end{equation}
\begin{equation} 
\label{eq23}
\frac{h^{''}}{h} - \frac{f^\prime h^\prime}{fh} = K,
\end{equation}
where prime and dot stand for differentiation with respect to $x$ and $t$ respectively.
Integrating (\ref{eq21}) leads to
\[
f = L h^{-\alpha},
\]
\begin{equation} 
\label{eq24}
g = M_{1} k^{\frac{n - n\alpha +1}{n - n\alpha + \alpha}},
\end{equation}
where $L$ and $M_{1}$ are constants of integration.\\
Here we consider three cases according to the values of $K$. \\
\par
\subsection {Case (1): $K > 0$}
Using (\ref{eq24}) in Equations (\ref{eq22}) and (\ref{eq23}) and after making 
suitable transformation of co-ordinates, the geometry of the spacetime (\ref{eq1})
 reduces to the form
\[
ds^{2} = a^{2} \cosh^{\frac{2n(\alpha - 1)}{N}}(K_{1}T)(dX^{2} 
- dT^{2}) \]
\[
+  \cosh^{-\frac{2\alpha}{\alpha + 1}}(K_{2}X) \cosh^{\frac{N +\alpha - 1}{N}} 
(K_{1}T)dY^{2} \]
\begin{equation} 
\label{eq25}
+ \cosh^{\frac{2}{\alpha + 1}}(K_{2}X) \cosh^{\frac{N - \alpha + 1}
{N}} (K_{1}T)dZ^{2},
\end{equation}
where \\
\begin{eqnarray*}
K_{1} = \sqrt{KN}, \\
K_{2} = \sqrt{K(\alpha + 1)}, \\
N = 2n\alpha - 2n - \alpha - 1, \\
a = L M_{1} a_{2}^{\frac{n(1 - \alpha)}{n - n\alpha + \alpha}}.
\end{eqnarray*}
The effective pressure and density for the model (\ref{eq25}) are given by
\[
8\pi \bar{p} = 8\pi(p - \xi \theta) = \frac{1}{a^{2}\cosh^{\frac{2n(\alpha - 1)}
{N}} (K_{1}T)}\times \]
\begin{equation} 
\label{eq26}
\left((n - n\alpha - \alpha + 2)K + \frac{\alpha^{2}K}{\alpha + 1}\tanh^{2}(K_{2}X) +
K_{3}\frac{K^{2}\tanh^{2}(K_{1}T)}{K^{2}_{1}}\right) - \Lambda,
\end{equation}
\[
8\pi \rho = \frac{1}{a^{2}\cosh^{\frac{2n(\alpha - 1)}{N}} 
(K_{1}T)}\times \]
\begin{equation} 
\label{eq27}
\left((n\alpha - n - \alpha)K - \frac{\alpha(\alpha - 2)K}{\alpha + 1}\tanh^{2}(K_{2}X) +
K_{3}\frac{K^{2}\tanh^{2}(K_{1}T)}{K^{2}_{1}}\right) + \Lambda,
\end{equation}
where
\[
K_{3} = n^{2}(3\alpha^{2} - 4\alpha + 2) - 2n(\alpha - 1)(\alpha - 3) -2 (\alpha^{2} - \alpha -1).
\]
Here $\theta$ is the scalar of expansion calculated for the flow vector $v^{i}$ as 
\begin{equation} 
\label{eq28}
\theta = \frac{N_{1}K\tanh(K_{1}T)}{a K_{1}\cosh^{\frac{n(\alpha - 1)}
{N}} (K_{1}T)},
\end{equation}
where
\[
N_{1} = 3n\alpha - 3n -\alpha - 1
\]
For the specification of $\xi$, we assume that the fluid obeys an equation of state of the form
\begin{equation} 
\label{eq29}
p = \gamma \rho,
\end{equation}
where $\gamma(0 \leq \gamma \leq 1)$ is a constant.\\
Thus, given $\xi(t)$ we can solve for the cosmological parameters. In most of the 
investigations involving bulk viscosity is assumed to be a simple power function of 
the energy density.\cite{ref85}$^-$\cite{ref87} 
\begin{equation}
\label{eq30}
\xi(t) = \xi_{0} \rho^{m},
\end{equation}
where $\xi_{0}$ and $m$ are constants. If $m = 1$, Eq. (\ref{eq30}) may correspond
to a radiative fluid.\cite{ref88} However, more realistic models\cite{ref89} are 
based on $m$ lying in the regime $0 \leq m \leq \frac{1}{2}$. \\
On using (\ref{eq30}) in (\ref{eq26}), we obtain
\[
8\pi(p - \xi_{0}\rho^{m} \theta) = \frac{1}{a^{2}\cosh^{\frac{2n(\alpha - 1)}
{N}} (K_{1}T)}\times \]
\begin{equation} 
\label{eq31}
\left((n - n\alpha - \alpha + 2)K + \frac{\alpha^{2}K}{\alpha + 1}\tanh^{2}(K_{2}X) +
K_{3}\frac{K^{2}\tanh^{2}(K_{1}T)}{K^{2}_{1}}\right) - \Lambda,
\end{equation} 
\subsubsection {Model I: ~ ~ ~ $(\xi = \xi_{0})$}
When $m = 0$, Equation (\ref{eq30}) reduces to $\xi = \xi_{0}$. With the use of 
Equations (\ref{eq27}), (\ref{eq28}) and (\ref{eq29}), Equation (\ref{eq31}) reduces to
\[
8\pi (1 + \gamma) \rho = \frac{8\pi N_{1}\xi_{0} K\tanh(K_{1}T)}
{aK_{1}\cosh^{\frac{n(\alpha -1)}{N}}(K_{1}T)} +
\]
\begin{equation} 
\label{eq32}
\frac{2(1 - \alpha)(\alpha + 1)K K^{2}_{1} + 2\alpha K K^{2}_{1} 
\tanh^{2}(K_{2}X) + 2(\alpha + 1)K^{2}K_{3}\tanh^{2}(K_{1}T)}{(\alpha + 1)a^{2}K^{2}_{1} 
\cosh^{\frac{2n(\alpha -1)}{N}}(K_{1}T)}.
\end{equation} 
Eliminating $\rho(t)$ between (\ref{eq27}) and (\ref{eq32}), we get
\[
(1 + \gamma)\Lambda = \frac{8\pi N_{1}\xi_{0} K\tanh(K_{1}T)}
{aK_{1}\cosh^{\frac{n(\alpha -1)}{N}}(K_{1}T)} +
\]
\[
\frac{1}{(\alpha + 1)a^{2}K^{2}_{1}\cosh^{\frac{2n(\alpha -1)}{N}}
(K_{1}T)}\times
\]
\[
\left[\{n - n\alpha - \alpha + 2 - (n\alpha - n - \alpha)\gamma\} (\alpha + 1)
K K^{2}_{1} +
\right.
\]
\[ 
\{ \alpha + (\alpha - 2)\gamma\} \alpha K K^{2}_{1}\tanh^{2}(K_{2}X) 
\]
\begin{equation} 
\label{eq33}
 \left. +  (1-\gamma)(\alpha + 1)K^{2}K^{2}_{3}\tanh^{2}(K_{1}T)\right]
\end{equation}
\subsubsection {Model II: ~ ~ ~ $(\xi = \xi_{0}\rho)$}
When $m = 1$, Equation (\ref{eq30}) reduces to $\xi = \xi_{0} \rho$. With the use of 
(\ref{eq27}), (\ref{eq28}) and (\ref{eq29}), Equation (\ref{eq31}) reduces to
\[
8\pi\rho\left[1 + \gamma - \frac{N_{1}\xi_{0} K\tanh(K_{1}T)}
{aK_{1}\cosh^{\frac{n(\alpha -1)}{N}}(K_{1}T)}\right] =
\]
\begin{equation} 
\label{eq34}
\frac{2(1 - {\alpha}^2) K K^{2}_{1} + 2 \alpha K K^{2}_{1} 
\tanh^{2}(K_{2}X) + 2(\alpha + 1)K^{2} K_{3}\tanh^{2}(K_{1}T)}{(\alpha + 1)
a^{2}K^{2}_{1} \cosh^{\frac{2n(\alpha -1)}{N}}(K_{1}T)}.
\end{equation}
Eliminating $\rho(t)$ between (\ref{eq27}) and (\ref{eq34}), we get
\[
\Lambda \left[1 + \gamma - \frac{N_{1} \xi_{0} K\tanh(K_{1}T)}{aK_{1}\cosh^{\frac
{n(\alpha - 1)} {N}} (K_{1}T)}\right] = \]
\[
\frac{(n - n\alpha - \alpha + 2)(\alpha + 1) K K^{2}_{1} + \alpha^{2} K K^{2}_{1} 
\tanh^{2}(K_{2}X) + (\alpha + 1)K^{2} {K_3}\tanh^{2}(K_{1}T)}{(\alpha + 1)a^{2}
K^{2}_{1}\cosh^{\frac{2n(\alpha -1)}{N}}(K_{1}T)}~ -
\]
\[
\frac{(n\alpha - n - \alpha)(\alpha + 1)K K^{2}_{1} - \alpha(\alpha - 2)K K^{2}_{1}\tanh^{2}
(K_{2}X) + (\alpha + 1)K^{2}K_{3}\tanh^{2}(K_{1}T)}{(\alpha + 1)a^{2}K^{2}_{1} 
\cosh^{\frac{2n(\alpha -1)}{N}}(K_{1}T)}\times
\]
\begin{equation} 
\label{eq35}
\left[\gamma - \frac{N_{1}\xi_{0} K\tanh(K_{1}T)}
{aK_{1}\cosh^{\frac{n(\alpha -1)}{N}}(K_{1}T)}\right]
\end{equation}
From Equations (\ref{eq33}) and (\ref{eq35}), we observe that the cosmological
constant is a decreasing function of time and it approaches a small value as
time progresses (i.e., the present epoch), which explains the small value of 
$\Lambda$ at present.  Figure 1 clearly shows this behaviour of $\Lambda$ 
as decreasing function of time in both the models I and II.\\
\begin{figure}[ht] 
\vspace*{13pt}
\centerline{\psfig{file=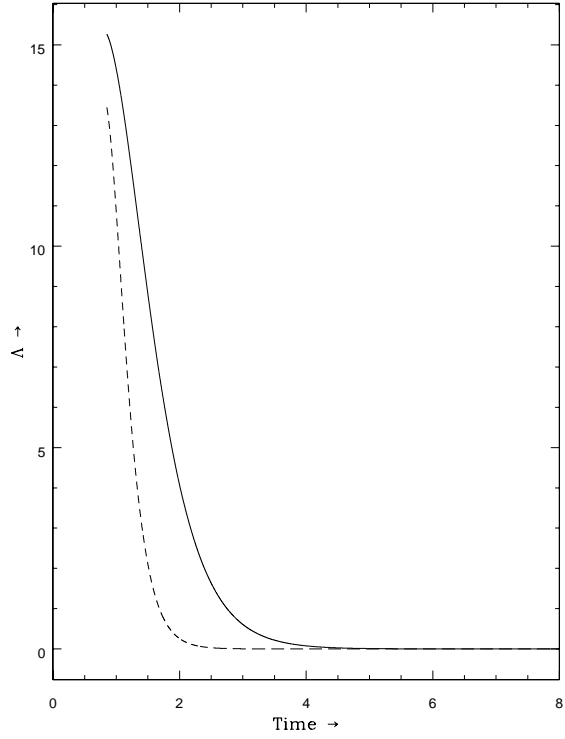,width=8cm}} 
\vspace*{13pt}
\caption{The plot of cosmological constant ($\Lambda$) with time for model 
$K > 0$ with parameters $\alpha = 1.5,  ~ \gamma = 0.5, ~  K = 1,~ a = 1, 
~ {\xi}_{0} = 1$, solid line for power index $m = 0$, and dashed line for 
$m = 1$  respectively.}
\end{figure}
 
{\bf Some Physical Aspects  of the Models} \\
We shall now give the expressions for kinematical quantities and the components
of conformal curvature tensor. With regard to the kinematical properties of the 
velocity vector $v^{i}$ in the metric (\ref{eq25}), a straight forward calculation 
leads to the following expressions for the shear of the fluid.
\begin{equation} 
\label{eq36}
\sigma_{11} = \frac{a(\alpha + 1)K}{3K_{1}}\cosh^{\frac{n(\alpha - 1)}
{N}}(K_{1}T)\tanh(K_{1}T),
\end{equation}
\begin{equation} 
\label{eq37}
\sigma_{22} = \frac{(\alpha - 2)K}{3aK_{1}}\cosh^{\frac{-2\alpha}{\alpha + 1}}(K_{2}X)
\cosh^{\frac{n\alpha - n - 2}{N}}(K_{1}T)\tanh(K_{1}T),
\end{equation}
\begin{equation} 
\label{eq38}
\sigma_{33} = \frac{(1 - 2\alpha)\sqrt{K}}{3aK_{1}}\cosh^{\frac{2}{\alpha + 1}}(K_{2}X)
\cosh^{\frac{n\alpha -  n - 2\alpha }{N}}(K_{1}T)\tanh(K_{1}T),
\end{equation}
\begin{equation} 
\label{eq39}
\sigma_{44} = \frac{a N_{1}K}{3K_{1}}\cosh^{\frac{n(\alpha - 1)}
{N}}(K_{1}T)\tanh(K_{1}T),
\end{equation}
where all other components, the rotation tensor $\omega$ and acceleration vanish.
The expansion scalar $\theta$ has already been given by (\ref{eq49}). Since 
$\dot{v_{i}} = v_{i \, ;j} v^{j} = 0$, the motion is geodetic.\\
The non-vanishing physical components of conformal curvature tensor are given by
\[
C^{12}_{12} = C^{34}_{34} = \frac{1}{6a^{2}\cosh^{\frac{2n(\alpha - 1)}
{N}}(K_{1}T)}\left[4\alpha K\tanh^{2}(K_{2}X) - (\alpha + 1)K + \right.
\]
\begin{equation} 
\label{eq40}
 \left. \{ n^{2}(\alpha - 1)^{2} - n (5\alpha^{2} - 4 \alpha + 1) +
 2\alpha \}\frac{K^{2}\tanh^{2}(K_{1}T)}
{K^{2}_{1}}\right] 
\end{equation}
\[
C^{13}_{13} = C^{24}_{24} = \frac{1}{6a^{2}\cosh^{\frac{2n(\alpha - 1)}
{N}}(K_{1}T)}\left[4K - 2\alpha K\tanh^{2}(K_{2}X) + \right.
\] 
\begin{equation} 
\label{eq41}
 \left. \{ 2n(\alpha - 2)(\alpha - 1) + 2\alpha \} \frac{K^{2}\tanh^{2}(K_{1}T)}{K^{2}_{1}}\right]  
\end{equation}
\[
C^{14}_{14} = C^{23}_{23} = \frac{1}{6a^{2} \cosh^{\frac{2n(\alpha - 1)}
{N}}(K_{1}T)} \left[ \frac{2(n\alpha^{2} - 2\alpha -1)K^{2}}{K^{2}_{1}}
\tanh^{2}(K_{1}T) - \right.
\]
\begin{equation} 
\label{eq42} 
 \left. 2\alpha K \tanh^{2}(K_{2}X) -2K \right]
\end{equation}
\begin{equation} 
\label{eq43}
C^{24}_{12} = C^{34}_{13} = \frac{n(1 - \alpha)\sqrt{(\alpha + 1)} 
K^{\frac{3}{2}}\tanh(K_{1}T)
\tanh(K_{2}X)}{2a^{2}K_{1}\cosh^{\frac{2n(\alpha - 1)}{N}}(K_{1}T)}
\end{equation}
The non-vanishing component $F_{12}$ of electro-magnetic field tensor and
$J^{2}$, the component of charge current density, are obtained as
\begin{equation} 
\label{eq44}
F^{2}_{12} = \left(\frac{\bar{\mu}\alpha K}{4\pi}\right)~ {\rm sech}^{\frac{2(2\alpha + 1)}
{\alpha + 1}} (K_{2}X) \cosh^{\frac{N + \alpha - 1}{N}}(K_{1}T)
\end{equation}
\[
J^{2} = \frac{1}{4a^{2}}\sqrt{\frac{\alpha K}{\bar{\mu}\pi(\alpha + 1)}}\cosh^{-\frac{1}
{\alpha + 1} } (K_{2}X) ~ {\rm  sech }^{\frac{N_{1} + \alpha}{N}}(K_{1}T)
\]
\begin{equation} 
\label{eq45}
\left[ {\bar{\mu}}^{\prime}\sqrt{(\alpha + 1)} - 
2 \alpha \sqrt{K}\bar{\mu}\tanh(K_{2}X)\right]
\end{equation}
The source of electro-magnetic field exist, when matter and charge are present and
does not exist when matter is absent.
The models represent expanding, shearing, non-rotating and Petrov type I non-degenerate
in general, in which the flow is geodetic. The model starts expanding at $T > 0$ and goes 
on expanding indefinitely. It is observed that the expansion is minimum at $T = 0$.  
Since $\lim_{T \rightarrow \infty}~\frac{\sigma}{\theta} \ne 0$,
the models do  not approach isotropy for large values of $T$. The pressure contrast 
$\frac{p_{x}}{p}$ and density contrast $\frac{\rho_{x}}{\rho}$ tends to zero for large 
values of $T$ which shows that inhomogeneity dies out for large values of $T$. 
When $\alpha = - 1$ the model reduces to its homogeneous form. 
It is remarkable to mention here that at the 
time of minimum expansion matter density dominates over the expansion in the model whereas
at the time of maximum expansion, expansion in the model dominates over matter density.\\ 
\par
\subsection {Case (2) : $K < 0$,~~{\rm  let} $K = -M$, $M > 0$}
Using (\ref{eq24}) in Equations (\ref{eq22}) and (\ref{eq23}) and after making 
suitable transformation of co-ordinates, the geometry of the spacetime (\ref{eq1})
 reduces to the form
\[
ds^{2} = b^{2}\cos^{\frac{2n(\alpha - 1)}{N}} (K_{4}T)(dX^{2} 
- dT^{2}) \]
\[
+\cos^{-\frac{2\alpha}{\alpha + 1}}(K_{5}X) \cos^{-\frac{(N + \alpha - 1))}{N}} 
(K_{4}T)dY^{2} \]
\begin{equation} 
\label{eq46}
+ \cos^{\frac{2}{\alpha + 1}}(K_{5}X)\cos^{\frac{-(N + + \alpha - 1)}
{N}} (K_{4}T)dZ^{2},
\end{equation}
where \\
\begin{eqnarray*}
K_{4} = \sqrt{MN}, \\
K_{5} = \sqrt{M(\alpha + 1)}, \\
b = L M_{1} {a^{\prime}_{2}}^{\frac{n(1 - \alpha)}{n - n\alpha + \alpha}}
\end{eqnarray*}
The effective pressure and density for the model (\ref{eq46}) are given by
\[
8\pi \bar{p} = 8\pi(p - \xi \theta) = \frac{1}{b^{2}\cos^{\frac{2n(1 - \alpha)}
{N}} (K_{4}T)}\times \]
\begin{equation} 
\label{eq47}
\left[-(N + 1)M + \frac{2\alpha^{2}M}{\alpha + 1}\tan^{2}(K_{5}X) +
K_{6}\frac{M^{2} \tan^{2}(K_{4}T)}{K^{2}_{4}}\right] - \Lambda,
\end{equation}
\[
8\pi \rho = \frac{1}{b^{2}\cos^{\frac{2n(1 - \alpha)}{N}} 
(K_{4}T)}\times \]
\begin{equation} 
\label{eq48}
\left[(2 - \alpha)M - \frac{\alpha(\alpha - 2)M}{\alpha + 1}\tan^{2}(K_{5}X) +
K_{7}\frac{M^{2}\tan^{2}(K_{4}T)}{K^{2}_{4}}\right] + \Lambda,
\end{equation}
where
\[
K_{6} = n(\alpha - 1)(5n - 5n\alpha + 4\alpha + 4) - \alpha^{2} - \alpha -1,
\]
\[
K_{7} = n(\alpha - 1)(3n\alpha - 3n - 4\alpha) + \alpha^{2} + \alpha -1.
\]
and $\theta$ is the scalar of expansion calculated for the flow vector $v^{i}$ as 
\begin{equation} 
\label{eq49}
\theta = \frac{N_{1}M\tan(K_{4}T)}{b K_{4}\cos^{\frac{n(1 - \alpha)}
{N}} (K_{4}T)}
\end{equation}
\par
\subsubsection {Model I: ~ ~ ~ $(\xi = \xi_{0})$}
When $m = 0$, Equation (\ref{eq30}) reduces to $\xi = \xi_{0}$. In this case 
Equation (\ref{eq47}) with the use of  (\ref{eq29}), (\ref{eq48}) and (\ref{eq49})
reduces to
\[
8\pi (1 + \gamma) \rho = \frac{8\pi N_{1}\xi_{0} M\tan(K_{4}T)}
{bK_{4}\cos^{\frac{n(1 - \alpha)}{N}}(K_{4}T)}~ +
\]
\begin{equation} 
\label{eq50}
\frac{-(N + 1)(\alpha + 1)M K^{2}_{4} + \alpha (\alpha + 2) M K^{2}_{4} 
\tan^{2}(K_{5}X) + (K_{6} + K_{7})(\alpha + 1) M^{2}\tan^{2}(K_{4}T)}{(\alpha + 1)b^{2}
K^{2}_{4} \cos^{\frac{2n(1 - \alpha)}{N}}(K_{4}T)}.
\end{equation} 
Eliminating $\rho(t)$ between (\ref{eq48}) and (\ref{eq50}), we get
\[
(1 + \gamma)\Lambda = \frac{8\pi N_{1}\xi_{0} M\tan(K_{4}T)}
{bK_{4} \cosh^{\frac{n(1 - \alpha)}{N}}(K_{4}T)} +
\]
\[
\frac{1}{(\alpha + 1)b^{2}K^{2}_{4}\cos^{\frac{2n(1 - \alpha)}{N}}
(K_{4}T)}\times
\]
\[
\left[\{2(n -\gamma) - \alpha(2n - \gamma) + \alpha\}(\alpha + 1)M K^{2}_{4}
+ 
\right.
\]
\[ \{\alpha^{2}(2 + \gamma) - 2\alpha \gamma \}M K^{2}_{4}\tan^{2}(K_{5}X) 
\]
\begin{equation} 
\label{eq51}
 \left. +  (K_{6} - \gamma K_{7})(\alpha + 1)M^{2}\tan^{2}(K_{4}T)\right]
\end{equation}
\par
\subsubsection {Model II: ~ ~ ~ $(\xi = \xi_{0}\rho)$}
When $m = 1$, Equation (\ref{eq30}) reduces to $\xi = \xi_{0} \rho$. In this case
Equation (\ref{eq47}) with the use of (\ref{eq29}), (\ref{eq48}) and (\ref{eq49}),
reduces to
\[
8\pi\rho\left[1 + \gamma - \frac{N_{1}\xi_{0} M \tan(K_{4}T)}
{bK_{4}\cos^{\frac{n(1 - \alpha)}{N}}(K_{4}T)}\right] =
\]
\begin{equation} 
\label{eq52}
\frac{-2(n + 1)({\alpha}^2 - 1)M K^{2}_{4} + \alpha(\alpha + 2) M K^{2}_{4} 
\tan^{2}(K_{5}X) + (K_{6} + K_{7})(\alpha + 1)M^{2}\tan^{2}(K_{4}T)}{(\alpha + 1)
b^{2}K^{2}_{4} \cos^{\frac{2n(1 - \alpha)}{N}}(K_{4}T)}.
\end{equation}
Eliminating $\rho(t)$ between (\ref{eq48}) and (\ref{eq52}), we get
\[
\Lambda \left[1 + \gamma - \frac{N_{1} \xi_{0} M \tan(K_{4}T)}{bK_{4}\cos^{\frac{n(1 - \alpha)}{N}} (K_{4}T)}\right] = \]
\[
\frac{(2n - 2n\alpha - \alpha)(\alpha + 1)M K^{2}_{4} + 2\alpha^{2}M K^{2}_{4} 
\tan^{2}(K_{5}X) + (\alpha + 1)K_{6}M^{2}\tan^{2}(K_{4}T)}{(\alpha + 1)b^{2}K^{2}_{4}
\cos^{\frac{2n(1 - \alpha)}{N}}(K_{4}T)} -
\]
\[
\frac{(2 - \alpha)(\alpha + 1)M K^{2}_{4} - \alpha(\alpha - 2)M K^{2}_{4}\tan^{2}
(K_{5}X) + (\alpha + 1)M^{2}K_{7}\tan^{2}(K_{4}T)}{(\alpha + 1)b^{2}K^{2}_{4} 
\cos^{\frac{2n(1 - \alpha)}{N}}(K_{4}T)}\times
\]
\begin{equation} 
\label{eq53}
\left[\gamma - \frac{N_{1}\xi_{0} M\tan(K_{4}T)}
{bK_{4}\cos^{\frac{n(1 - \alpha)}{N}}(K_{4}T)}\right]
\end{equation}
Our study for the case $ K < 0 $ shows constant value of cosmological constant
($\Lambda$) for large values of time and do not decrease with time (this means that 
the universe in not expanding or may be  steady state condition). In this case,
detailed study shows that the scalar of expansion $\theta$ does not increase 
with time. Our study is inconsistent with work done by Singh {\it et al.}
\cite{ref82} The Singh {\it et al.} claim that the universe is expanding which 
does not match with our result. Also the claim of minimum and maximum expansion
rate in $\theta$ is reflection of periodicity of trigonometric functions 
involved there. We are trying to find feasible interpretation and situations 
relevant to $ K < 0 $. Further study is in progress.  \\

\noindent
{\bf Some Physical Aspects of the Models} \\
The non-vanishing components of shear tensor $\sigma_{ij}$ are obtained as 
\begin{equation} 
\label{eq54}
\sigma_{11} = \frac{b(\alpha + 1)M}{3K_{4}}~{\rm sec}^{\frac{n(\alpha - 1)}
{N}}(K_{4}T)\tan(K_{4}T),
\end{equation}
\begin{equation} 
\label{eq55}
\sigma_{22} = \frac{(\alpha - 2)M}{3bK_{4}}~{\rm sec}^{\frac{2\alpha}{\alpha + 1}}(K_{5}X)
~{\rm sec}^{\frac{n\alpha - n - 2}{N}}(K_{4}T)\tan(K_{4}T),
\end{equation}
\begin{equation} 
\label{eq56}
\sigma_{33} = \frac{(1 - 2\alpha)M}{3bK_{4}}~{\rm sec}^{-\frac{2}{\alpha + 1}}(K_{5}X)
~{\rm sec}^{\frac{n\alpha -  n - 2\alpha }{N}}(K_{4}T)\tan(K_{4}T),
\end{equation}
\begin{equation} 
\label{eq57}
\sigma_{44} = \frac{N_{1}M}{3bK_{4}}~{\rm sec}^{\frac{n(\alpha - 1)}
{N}}(K_{4}T)\tan(K_{4}T)
\end{equation}
The expansion scalar $\theta$ has already been given by (\ref{eq49}).
The rotation $\omega$ is identically zero. Since $\dot{v_{i}} = v_{i \, ;j} V^{j} = 0$, the 
motion is geodetic.\\
The non-vanishing physical components of conformal curvature tensor are given by
\[
C^{12}_{12} = C^{34}_{34} = \frac{1}{6b^{2}\cos^{\frac{2n(1 - \alpha)}
{N}}(K_{4}T)}\left[(4\alpha + n - n\alpha)M + 4\alpha M\tan^{2}(K_{5}X) +  \right.
\]
\begin{equation} 
\label{eq58}
 \left. \{ n(\alpha - 1)(3n - 3n\alpha + 5 \alpha - 1) - 4\alpha^{2} + 2\}
\frac{M\tan^{2}(K_{4}T)}{N}\right] 
\end{equation}
\[
C^{13}_{13} = C^{24}_{24} = \frac{1}{6b^{2}\cos^{\frac{2n(1 - \alpha)}
{N}}(K_{4}T)}\left[\frac{ \{2n(1 - \alpha)(\alpha - 2) + \alpha^{2} - 2 \}}
{K^{2}_{4}} M^{2}\tan^{2}(K_{4}T)  \right.
\] 
\begin{equation} 
\label{eq59}
 \left. - 2\alpha M {\rm sec^{2}(K_{5}X)}\right]  
\end{equation}
\[
C^{14}_{14} = C^{23}_{23} = \frac{1}{6b^{2} \cos^{\frac{2n(1 - \alpha)}
{N}}(K_{4}T)} \left[ \frac{2(\alpha^{2} - n\alpha^{2}  + n + 1)}
{K^{2}_{4}} M^{2} \tan^{2}(K_{4} T)  \right.
\]
\begin{equation} 
\label{eq60} 
 \left. - 2\alpha M~ {\rm sec}^{2}(K_{5}X) \right]
\end{equation}
\begin{equation} 
\label{eq61}
C^{24}_{12} = C^{34}_{13} = \frac{n(\alpha - 1)\sqrt{(\alpha + 1)} 
M^{\frac{3}{2}}\tan(K_{4}T)
\tan(K_{5}X)}{2 b^{2} K_{4}\cos^{\frac{2n(1 - \alpha)}{N}}(K_{4}T)}
\end{equation}
The non-vanishing component $F_{12}$ of electro-magnetic field tensor and
$J^{2}$, the component of charge current density, are obtained as
\[
F^{2}_{12} = \left(\frac{\bar{\mu}M}{4\pi}\right)~{\rm sec}^{\frac{2\alpha}
{\alpha + 1}} (K_{5}X) {\rm sec}^{\frac{N + \alpha - 1}{N}}(K_{4}T)
\]
\begin{equation} 
\label{eq62}
\left[\frac{\{ \alpha^{2}(2n - 1) - 2n(2\alpha - 1) + 1\}}{K^{2}_{4}} M\tan^{2}(K_{4}T)
- \alpha\tan^{2}(K_{5}X) - 1\right]
\end{equation}
\[
J^{2} = \frac{1}{4b^{2}}\sqrt{\frac{M}{\bar{\mu}\pi H_{1}(\alpha + 1)}} ~ {\rm sech }^
{\frac{-(N_{1} + \alpha)}{N}}(K_{4}T)
\]
\[
{\rm sec}^{\frac{-\alpha}{\alpha + 1} }(K_{5}X)\left[2\bar {\mu} \alpha(\alpha + 1)
\sqrt{M}\tan(K_{5}X) \right.
\]
\begin{equation} 
\label{eq63}
\left.  {\rm sec}^{2}(K_{5}X) - \bar{\mu}^{\prime}H_{1}\sqrt{(\alpha + 1)} + 2\bar{\mu}
H_{1}\sqrt{M}\tan(K_{5}X)\right]
\end{equation}
where
\[
H_{1} = \left[\frac{\{\alpha^{2}(2n - 1) - 2n(2\alpha - 1) + 1\}}{K^{2}_{4}} 
M^{2}\tan^{2}(K_{4}T) - \alpha\tan^{2}(K_{5}X) - 1\right]
\]
The models represent  shearing, non-rotating and Petrol type I non-degenerate
in general, in which the flow is geodetic. The model starts expanding at $T > 0$
but the initial expansion is slow. When $T$ is closer to $\frac{\pi}{2K_{4}}$,
it has stiff rise in the expansion then decreases. This shows the case of 
oscillation. It is observed that the expansion is minimum at $T = 0$ or 
$T = \frac{\pi}{K_{4}}$. The large values of $\theta$ {\rm near} 
$ T = \frac{\pi}{2 K_{4}}$ is reflection of trigonometric property. But expansion
remains finite.  Since $\lim_{T \rightarrow \infty}~\frac{\sigma}{\theta}
 \ne 0$, hence the models do  not approach isotropy for large values of $T$. 
In this case inhomogeneity also dies out for large value of $T$. \\ 
\par
\subsection {Case (3): $K = 0$}
Using (\ref{eq24}) in Equations (\ref{eq22}) and (\ref{eq23}) and after making 
suitable transformation of co-ordinates, the geometry of the spacetime (\ref{eq1})
reduces to the form
\begin{equation} 
\label{eq64}
ds^{2} = c^{2} T^{\frac{2n(\alpha - 1)}{N}}(dX^{2} - dT^{2}) + X^{\frac{-2\alpha}
{\alpha +1}} T^{\frac{N + \alpha - 1}{N}} dY^{2} + X^{\frac{2}{\alpha + 1}}
T^{\frac{N - \alpha + 1}{N}} dZ^{2}, 
\end{equation}
where
\[
c = LM_{1} {a_{2}^{\prime \prime } }^{\frac{n(1 - \alpha)}{n - n\alpha + \alpha}}.
\]
It is observed that the model (\ref{eq64}) starts to expand from its singularity
stage i.e. at $T = 0$ and goes expanding indefinitely when ${T \rightarrow \infty}$.
The model represents expanding, shearing, non-rotating and Petrov type I non-
degenerate in general in which the flow is geodetic. This model also does not approach
isotropy for large values of $T$. It is also observed here at the time of initial
singularity the matter density dominates over the expansion of the model. \\
\par
\section{Conclusions}
We have obtained a new class of cylindrically symmetric inhomogeneous cosmological
models electro-magnetic bulk viscous fluid as the source of matter. Generally the 
models represent expanding, shearing, non-rotating and Petrov type-I non-degenerate
in which the flow vector is geodetic. In all these models, we observe that they do 
not approach isotropy for large values of time. It is concluded that $(i)$ if 
 $\alpha > 0$ then for the cases $K > 0$ or the values of $K$ tending to $0$, 
the models have point type singularity
at the time of maximum expansion and $(ii)$ if  $0 < \alpha < 1$ then they
have infinite singularity. Whereas the case $K < 0 $ is just opposite of the case
$K \geq 0$. In all the cases the spacetime is conformally flat for large values of
$T$ and at the time of minimum expansion matter density dominates over expansion.
For $K = 0 $ the material energy is more dominant over magnetic energy. \\
The cosmological constant in all models given in section 3 are decreasing 
function of time and they all approach a small value as time increases (i.e., the
present epoch) except the case $ K < 0 $ . The values of cosmological ``constant'' 
for these models are found to be small and positive which are supported by the 
results from recent supernovae observations recently obtained by 
the High - z Supernova Team and Supernova Cosmological Project ( Garnavich
 {\it et al.}\cite{ref30};  Perlmutter 
{\it et al.}\cite{ref27}; Riess {\it et al.}\cite{ref28}; Schmidt {\it et al.}
\cite{ref33}). Thus, with our approach, we obtain a physically relevant decay 
law for the cosmological constant unlike other investigators where {\it adhoc} 
laws were used to arrive at a mathematical expressions for the decaying vacuum 
energy. Thus our models are more general than those studied earlier.  
We would like to find  feasible situations for $ K < 0 $. Our strong point of this 
model is that it in-cooperates matter density naturally and so makes feasible 
model which can be in-cooperates the physical constraints.\\ 
\par    
\nonumsection{Acknowledgements} 
A. Pradhan and Kanti R. Jotania  thank to the Inter-University Centre for 
Astronomy and Astrophysics, Pune, India for providing  facility under Associateship 
Programme where part of this work was carried out. \\
\newline
\newline
\nonumsection{References}

\end{document}